# GENERALIZED APPROACH TO DESCRIPTION OF ENERGY DISTRIBUTION OF SPIN SYSTEM

**Boris Kryzhanovsky and Leonid Litinskii**

Center of Optical Neural Technologies
Scientific Research Institute for System Analysis RAS,
Moscow, Vavilova St., 44/2, 119333, Russia
kryzhanov@mail.ru,    litin@mail.ru

**Abstarct.** We examined energy spectrums of some particular systems of $N$ binary spins. It is shown that the configuration space can be divided into $N$ classes, and in the limit $N \to \infty$ the energy distributions in these classes can be approximated by the normal distributions. For each class we obtained the expressions for the first three moments of the energy distribution, including the case of presence of a nonzero inhomogeneous magnetic field. We also derived the expression for the variance of the quasienergy distribution in the local minimum. We present the results of computer simulations for the standard Ising model and the Sherrington-Kirkpatrick and Edwards-Anderson models of spin glass. Basing on these results, we justified the new method of the partition function calculation.

## 1. Introduction

In [1,2] we discussed briefly a new method of the calculation of the partition function for a system of $N$ binary spins $s_i = \{\pm 1\}$, $i = 1, ..., N$. The configuration vector $\mathbf{S} = (s_1, s_2, ..., s_N)$ and the energy $E$,

$$E \equiv E(\mathbf{S}, \mathbf{H}) = -\sum_{i=1}^{N}\sum_{j=1}^{N} T_{ij} s_i s_j - \sum_{i=1}^{N} H_i s_i , \qquad (1)$$

define the state of the system. In Eq.(1) $\mathbf{H} = (H_1, H_2, ..., H_N)$ is the magnetic field, $\mathbf{T} = (T_{ij})$ is a symmetric matrix with all the diagonal elements equal to zero ($T_{ii} = 0$).

The main idea of our method is very simple. It can be explained as follows. Let us examine the partition function $Z = \sum_{\mathbf{S}} \exp[-\beta E(\mathbf{S})]$, where the summation is carried over all the states $\mathbf{S}$ and $\beta$ is the reciprocal temperature. To calculate $Z$ we need to know the energy distribution in the space of states $\mathbf{S}$. However, in the general case it is not known. Our approach includes a decomposition of the set of all the states into $N$ classes, and inside each class, we approximate the energy distribution by the normal distribution with the known mean and the variance.

When decomposing the space into classes, we choose an arbitrary point of the space as an initial configuration $\mathbf{S}_0$. The set of configurations that differ from $\mathbf{S}_0$ by the $n$ opposite values of spins we call the class $\Omega_n$. In fact, $\Omega_n$ is the $n$-vicinity of the initial configuration. The class $\Omega_n$ consists of $L$ configurations, where $L = C_N^n$ is the number of combinations of $N$ elements taken $n$ at a time. All the configurations from $\Omega_n$ have the same relative magnetization $m = \mathbf{SS}_0^+ / N = 1 - 2n/N$. The set of all the $n$-vicinities $\Omega_n$, $n = 0, 1, 2, ..., N$, exhausts all the variety of configurations, and the partition function $Z$ can be written down as

$$Z = \sum_{n=0}^{N} \sum_{\mathbf{S} \in \Omega_n} \exp[-\beta \cdot E(\mathbf{S})]. \tag{2}$$

In the general case, we do not know how the energies of the states belonging to $\Omega_n$ are distributed. From our computer simulations we know only that this distribution is a single-humped one. Moreover, it fits the normal distribution reasonable well. Let us make it clear what we have in mind. Let $\bar{E}_n$ and $\sigma_n^2$ be the values of the mean and the variance of the distribution for the states from $\Omega_n$, respectively. We can estimate their values with the aid of random sampling. Then for large $N$ the curve $\exp[-\frac{1}{2}(E - \bar{E}_n)^2 / \sigma_n^2] / \sqrt{2\pi}\sigma_n$ is a good approximation for the experimental one, at least in the energy interval $\bar{E}_n - 3\sigma_n \leq E \leq \bar{E}_n + 3\sigma_n$. Consequently, if we obtain analytical expressions for $\bar{E}_n$ and $\sigma_n$ we can replace the summation in Eq.(2) over the states $\Omega_n$ by integration with respect to Gaussian measure. This allows us to reduce the right-hand side of Eq.(2) to the integral of the exponent with a large number $N$ in its argument. Such integrals are calculated by the aid of the standard Laplace method (method of steepest decent). First results concerning this approach were presented in [1,2], however due to the shortage of place the justification of this method was not published.

In the present paper, we describe the energy spectrum of a spin system and substantiate the method of calculation of its partition function (see [1,2]). The partition function calculation is one of the main problems of statistical physics. At first, they used partition functions as related to physical issues only. However, in the course of time it becomes clear that partition functions are also important in analysis of neural networks [3], in the combinatorial optimization problems [4], in the machine learning [5] and so on.

Papers [6,7] were first attempt to describe the shape of the energy hypersurface (1). The authors proposed to part the space of configuration states into microcanonical ensembles and obtained asymptotic expressions for the first two moments of the energy distribution. A more accurate expressions for the means and the variances of energy distributions in the microcanonical ensembles are given in papers [8-10]. These expressions were successfully used when deriving the minimization algorithms [8-15] and describing complex neuron systems [16-19].

The paper is organized as follows. In Section 2, we give the derivation of exact expressions for $\bar{E}_n$ and $\sigma_n$. In Section 3, we obtain estimates for the variance of quasienergies in the local minimum, which we need for application of our method to the models with random matrices $\mathbf{T}$. In Section 4, we present the results of computer simulations that show how well the energy distribution for the class $\Omega_n$ can be approximated by the normal distribution. We show graphs where we compare the results of our simulations with the normal curve. In Section 5, we analyze our experimental data and discuss the obtained results. The Appendix contains some cumbersome computational details.

## 2. Calculation of moments of distribution.

By $\mathbf{S}_0 = (s_1^{(0)}, s_2^{(0)}, ..., s_N^{(0)})$ we denote the initial configuration (in what follows we use the configuration in place of the configuration vector). The initial configuration $\mathbf{S}_0$ can be any random configuration as well as a specially chosen one. Its $n$-vicinity $\Omega_n$ consists of all the configurations that differ from $\mathbf{S}_0$ by the values of $n$ coordinates. We are interested in the distribution of energies of these states. Let us show how the mean and the variance of the energy distribution of the states from the class $\Omega_n$ can be calculated:

$$\bar{E}_n = \frac{1}{L} \sum_{\mathbf{S} \in \Omega_n} E(\mathbf{S}), \qquad \sigma_n^2 = \frac{1}{L} \sum_{\mathbf{S} \in \Omega_n} \left[ E(\mathbf{S}) - \bar{E}_n \right]^2, \tag{3}$$

where $L = C_N^n$. At first, we perform all the calculations for zero magnetic field ($\mathbf{H} = 0$), and then we generalize them to the case $\mathbf{H} \neq 0$. By $E_0 = E(\mathbf{S}_0, \mathbf{H} = 0)$ we denote the value of the energy for the initial configuration $\mathbf{S}_0$ in the absence of magnetic field:

$$E_0 = -\sum_{i=1}^{N} \sum_{j=1}^{N} T_{ij} s_i^{(0)} s_j^{(0)}. \tag{4}$$

Let $\mathbf{S}^{(k)} = (s_1^{(k)}, s_2^{(k)}, ..., s_N^{(k)})$ be one of the configurations from the class $\Omega_n$, where $k = 1, ..., L$. Components of this configuration can be expressed in terms of the components of the initial configuration $\mathbf{S}_0$: $s_i^{(k)} = s_i^{(0)} a_i^{(k)}$. Here $\mathbf{A}^{(k)} = (a_1^{(k)}, a_2^{(k)}, ..., a_N^{(k)})$ is a vector whose n components are equal to -1 and $N - n$ ones are equal to +1. Then the energy at the point $S^{(k)}$ takes the form:

$$E(\mathbf{S}^{(k)}) = -\sum_{i=1}^{N} \sum_{j=1}^{N} T_{ij} s_i^{(0)} s_j^{(0)} a_i^{(k)} a_j^{(k)} \chi_{ij}, \tag{5}$$

where $\chi_{ij}$ is the symmetric second-rank unit tensor: $\chi_{ij} = 1$ if $i \neq j$, $\chi_{ij} = 0$ if $i = j$. The set of vectors $\{\mathbf{A}^{(k)}\}$ define the class $\Omega_n$ completely and averaging over the class $\Omega_n$ reduces to summation over all the vectors $\mathbf{A}^{(k)}$, $k = 1, ..., L$. Consequently, substituting Eq.(5) in Eq.(3) we can calculate an arbitrary moment of the energy distribution in the class $\Omega_n$.

It is easy to note (see Appendix) that the problem of calculation of the moments of the energy distribution reduces to average over the value $a_{i_1}^{(k)} a_{i_2}^{(k)} ... a_{i_r}^{(k)} \chi_{i_1 i_2 ... i_r}$ of the class $\Omega_n$, where $\chi_{i_1 i_2 ... i_r}$ is the symmetric $r$-rank unit tensor, that is $\chi_{i_1 i_2 ... i_r} = 1$ when all the indices are different, $\chi_{i_1 i_2 ... i_r} = 0$ if at least two indices are the same. In other words, we reduce the problem to finding of the mean value of the product of $r$ components of the vector $\mathbf{A}^{(k)}$:

$$K_r = \frac{1}{L} \sum_{\Omega_n} a_{i_1}^{(k)} a_{i_2}^{(k)} ... a_{i_r}^{(k)} \chi_{i_1 i_2 ... i_r} \equiv \frac{1}{L} \sum_{k=1}^{L} a_{i_1}^{(k)} a_{i_2}^{(k)} ... a_{i_r}^{(k)} \chi_{i_1 i_2 ... i_r} \qquad (6)$$

Indeed, when substituting Eq.(6) in Eq.(3) and changing the order of summation we obtain the expression for the mean value of the energy:

$$\bar{E}_n = -\sum_{i=1}^{N} \sum_{j=1}^{N} T_{ij} s_i^{(0)} s_j^{(0)} \cdot \frac{1}{L} \sum_{k=1}^{L} a_i^{(k)} a_j^{(k)} \chi_{ij} \qquad (7)$$

From general reasons it is evident that the factor $K_2 = L^{-1} \sum a_i^{(k)} a_j^{(k)} \chi_{ij}$ in Eq.(7) does not depend on the indices $i$ and $j$, and we can take it outside the summation sign. Then Eq.(7) takes the form:

$$\bar{E}_n = K_2 E_0.$$

For calculation of the values $K_r$ ($r = 1,...,6$) and accurate expressions for the first three moments of distribution see Appendix. Here we present the expressions for the mean and the variance only:

$$\bar{E}_n = E_0 \frac{(N-2n)^2 - N}{N(N-1)},$$

$$\sigma_n^2 = \frac{8n(N-n)}{N(N-1)(N-2)(N-3)} \left[ A \left( \operatorname{Sp} \mathbf{T}^2 - \frac{E_0^2}{N(N-1)} \right) + NB\sigma_\lambda^2 \right]. \qquad (8)$$

In these equations $\operatorname{Sp} \mathbf{T}^2$ is the trace of the matrix $\mathbf{T}^2$, $\sigma_\lambda^2$ is the variance of the quasienergies $\lambda_i = h_i s_i^{(0)}$ distribution, where $h_i$ is the local field acting on the $i$-th spin in the initial configuration $\mathbf{S}_0$. The following notations are used:

$$A = 4(n-1)(N-n-1), \qquad B = 2[(N-2n)^2 - (N-2)]$$

$$\sigma_\lambda^2 = \frac{1}{N} \sum_{i=1}^{N} \lambda_i^2 - \left( \frac{1}{N} \sum_{i=1}^{N} \lambda_i \right)^2, \qquad \operatorname{Sp} \mathbf{T}^2 = \sum_i \sum_j T_{ij}^2 \qquad (9)$$

$$\lambda_i = h_i s_i^{(0)}, \qquad h_i = \sum_{j \neq i} T_{ij} s_j^{(0)}, \qquad E_0 = -\sum \lambda_i$$

In the asymptotic limit $N \to \infty$ Eqs.(8) become particularly simple:

$$\bar{E}_n = E_0 \cdot m^2, \quad \sigma_n^2 = 2 \text{Sp}\mathbf{T}^2 \cdot (1-m^2)\left[(1-m^2)(1-\varepsilon_0^2) + 2m^2\bar{\sigma}_\lambda^2\right] \quad (10)$$

where где $m = 1 - 2n/N$ is the (relative) magnetization and the following dimensionless quantities are introduced:

$$\varepsilon_0^2 = \frac{E_0^2}{N^2 \text{Sp}\mathbf{T}^2}, \quad \bar{\sigma}_\lambda^2 = \frac{N\sigma_\lambda^2}{\text{Sp}\mathbf{T}^2}. \quad (11)$$

In the presence of the magnetic field, the expressions for the mean and the variance are more complicated (see Appendix). In the asymptotic limit $N \to \infty$ these expressions take the form:

$$\bar{E}_n(\mathbf{H}) = m^2 E_0 - m\mathbf{HS}_0^+, \quad (12)$$

$$\sigma_n^2(\mathbf{H}) = \sigma_n^2 + (1-m^2)\left[\|\mathbf{H}\|^2 - \frac{1}{N}(\mathbf{HS}_0^+)^2 + 4m\left(\mathbf{HTS}_0^+ - \frac{E_0}{N}\mathbf{HS}_0^+\right)\right], \quad (13)$$

where $\mathbf{S}_0^+$ is the transposed vector $\mathbf{S}_0$, and the values of $E_0$ and $\sigma_n^2$ are defined by Eq.(4) and Eq.(10), respectively. In the specific case when all the components of the magnetic field $\mathbf{H} = (H_1, H_2, ..., H_N)$ are equal in absolute values, $|H_i| = H$, and the initial configuration $\mathbf{S}_0$ is chosen along the magnetic field, then $\mathbf{HS}_0^+ = HN$ and Eqs.(12), (13) become substantially simpler:

$$\bar{E}_n(\mathbf{H}) = E_0 \cdot m^2 - mHN,$$

$$\sigma_n^2(\mathbf{H}) = \sigma_n^2.$$

Note, all the moments of the energy distribution inside the class $\Omega_n$ are defined entirely by the moments of the distribution of quasienergies $\lambda_i = h_i s_i^{(0)}$ in the initial configuration $\mathbf{S}_0$. For example, the mean $\bar{E}_n$ is defined by the value of $E_0 = -\sum \lambda_i$ that it is proportional to the mean value of the quasienergies $\lambda_i$. The variance $\sigma_n^2$ includes the dependence on the variances of the quasienergy distribution $\sigma_\lambda^2$. The third moment of the energy distribution also includes the dependence on the mean value of the quantities $\lambda_i^3$, and so on (see Appendix).

For uniform spin systems, where the neighborhoods of all the spins are the same, the calculation of the moments simplifies significantly if the ground state of the system (the global minimum of the energy) is taken as the initial configuration $\mathbf{S}_0$. For the Ising model on the hypercube, the ground state is well known (for discussion see Section 4.1), that is we know the quantities $E_0$ and $\sigma_\lambda^2 = 0$. For a system with a random matrix $\mathbf{T}$ (for example, for the Sherrington-Kirkpatrick or Edwards-Anderson models) the question arises: how one can calculate the energy of ground state and the variance of the quasienergy distribution $\sigma_\lambda^2$? In the following section, we answer this question.

### 3. Quasienergy distribution in minimum

Let us examine a spin system with a connection matrix $\mathbf{T}$, whose elements may be regarded as independent random variables characterized by a distribution with the zero mean and the variance $\sigma_T^2$. We are interested in the distribution of quasienergies $\lambda_i = h_i s_i^{(0)}$, where $h_i$ is the local field (9) acting on the $i$-th spin. Note that the quantity $\lambda_i$ defines the energy related to $i$-th spin. Indeed, the energy of the state is equal to the sum of quasienergies $E = -\sum \lambda_i$ and consequently the mean value of the quasienergy is equal to $\bar{\lambda}_i = -E/N$.

It is easy to see that if the initial configuration $\mathbf{S}_0$ is selected randomly, the values $T_{ij} s_i^{(0)} s_j^{(0)}$ are also random variables. In this case, the quasienergies $\lambda_i$ being the sum of large number of random independent variables $T_{ij} s_i^{(0)} s_j^{(0)}$ are distributed according the normal low with the variance $\sigma_\lambda^2 = N \sigma_T^2$.

It is very different if the configuration $\mathbf{S}_0$ corresponds to the energy minimum. In what follows we analyze this case only. Now $T_{ij}$ and $s_i^{(0)} s_j^{(0)}$ are not independent variables since due to their correlation the minimum appears. However, this circumstance can be overcome. Following [20] from the matrix $\mathbf{T}$ we single out the term $\mathbf{T}_0$ responsible for the minimum appearance. For that, we express the matrix $\mathbf{T}$ in the form

$$\mathbf{T} = \mathbf{T}_0 + \mathbf{T}_1, \tag{14}$$

where

$$\mathbf{T}_0 = r_0 \sigma_T \mathbf{S}_0^+ \mathbf{S}_0, \quad \mathbf{T}_1 = \mathbf{T} + \mathbf{T}_0.$$

The weight $r_0$ is determined by the absence of correlations between the elements of the matrices $\mathbf{T}_0$ and $\mathbf{T}_1$. Calculating the covariation of the matrix elements and supposed it to be equal to zero, we obtain

$$r_0 = -\frac{E_0}{N(N-1)\sigma_T}. \tag{15}$$

With this choice of the weight $r_0$ from Eqs.(17-19) we obtain the relation

$$\mathbf{S}_0 \mathbf{T}_1 \mathbf{S}_0^+ = 0.$$

It shows that the contribution of $\mathbf{T}_1$ into the energy $E_0$ is strictly equal to zero, and $E_0 = -\mathbf{S}_0\mathbf{T}_0\mathbf{S}_0^+$. Thus the minimum of the configuration $\mathbf{S}_0$ is due to the influence of $\mathbf{T}_0$ only. Moreover, given the relation (15) the standards of the elements of the matrices $\mathbf{T}_0$ and $\mathbf{T}_1$ has the form:

$$\sigma_0 = r_0 \sigma_T, \qquad \sigma_1 = \sigma_T \sqrt{1-r_0^2}.$$

From here we obtain the relation $\sigma_T^2 = \sigma_0^2 + \sigma_1^2$ confirming that in Eq. (14) the matrix $\mathbf{T}$ is the sum of two uncorrelated random matrices $\mathbf{T}_0$ and $\mathbf{T}_1$.

Taking into account Eqs. (17-19) we present the quasienergy in the configuration $\mathbf{S}_0$ in the form

$$\lambda_i = \sqrt{N} \sigma_1 (\gamma + x_i), \tag{16}$$

where

$$\gamma = \frac{|E_0|}{N^{3/2} \sigma_1}, \quad x_i = \frac{1}{\sqrt{N}\sigma_1} \sum_{j \neq i} T_{1,ij} s_i^{(0)} s_j^{(0)}. \tag{17}$$

From our analysis it follows that the random variables $T_{1,ij}$ and $s_i^{(0)} s_j^{(0)}$ are uncorrelated ones and the variables $T_{1,ij} s_i^{(0)} s_j^{(0)}$ are distributed according a law with zero mean and the standard $\sigma_1$. This means that in the limit $N \gg 1$ we can assume the variables $x_i$ to be normally distributed with zero mean and the standard $\sigma_x = 1$. However, it has to be taken into account that in the minimum each spin is directed along its own local field, that is $\lambda_{i0} \geq 0$ and at that $\sum \lambda_{i0} = -E_0$. This means that dimensionless values $x_i$ satisfy the conditions $x_i \geq -\gamma$ and $\sum x_i = 0$ that change the moments of distribution of the variables $x_i$ radically.

With account of the aforesaid, the problem of determination of the distribution of quasienergies $\lambda_i$ is reduced to the following one. Suppose we have a set of random variables $x_i$. Taken separately each of them we can consider as normally distributed one with zero mean and unit variance. We are interested in their joint distribution when the random variables $x_i$ are cannot be considered as independent ones and the conditions $x_i \geq -\gamma$ and $\sum x_i = 0$ are true.

We write the joint distribution of the variables $x_i$ as

$$P_N(x_1, x_2, ..., x_N) = \frac{1}{P_0} \prod_{i=1}^{N} P(x_i) \, \delta(x_1 + x_2 + ... + x_N), \tag{18}$$

where $P_0$ is the normalization constant and

$$P(x) = \frac{1}{\sqrt{2\pi}} e^{-\frac{1}{2}x^2} \text{ when } x \geq -\gamma, \quad P(x) = 0 \text{ when } x < -\gamma.$$

At first, let us define the constant $P_0$. When integrating Eq.(18) over all $x_i \geq -\gamma$ and using the integral representation for the delta function, we obtain:

$$P_0 = \frac{1}{2\pi} \int_{-\infty}^{\infty} f_\omega^N d\omega, \qquad f_\omega = \frac{1}{\sqrt{2\pi}} \int_{-\gamma}^{\infty} e^{-\frac{1}{2}x^2 - i\omega x} dx. \tag{19}$$

We estimate the integral (19) by the saddle point method. From the condition $df_\omega/d\omega = 0$ we obtain that coordinate of the saddle point is $\omega = -i\delta$, where $\delta$ is the solution of the equation

$$\delta = \frac{1}{\sqrt{2\pi}\,\Phi_0} e^{-\frac{1}{2}(\gamma-\delta)^2}, \tag{20}$$

$$\Phi_0 = \frac{1}{\sqrt{2\pi}} \int_{-\gamma+\delta}^{\infty} e^{-\frac{1}{2}t^2} dt.$$

Then the constant $P_0$ takes the form:

$$P_0 = \frac{1}{\sqrt{2\pi N(1-\gamma\delta)}} \Phi_0^N e^{\frac{1}{2}N\delta^2}.$$

Now let us analyze the marginal distribution of the variables $x_i$. When integrating the left-hand side of Eq.(18) over all coordinates $x_j$ with $j \neq i$, we obtain omitting the subscript $i$:

$$\bar{P}(x) = \frac{1}{2\pi P_0} P(x) \int_{-\infty}^{\infty} f_\omega^{N-1} e^{-i\omega x} d\omega.$$

The estimate of this integral with the aid the saddle point method leads to the following expression:

$$\bar{P}(x) = \frac{1}{\sqrt{2\pi}\,\Phi_0} e^{-\frac{1}{2}(x-\delta)^2} \text{ for } x \geq -\gamma, \qquad \bar{P}(x) = 0 \text{ for } x < -\gamma. \tag{21}$$

It is easy to see that this distribution satisfies the normalization conditions and the mean and the variance of the marginal distribution (21) are defined by the expressions:

$$\bar{x} = 0, \qquad \sigma_x^2 = 1 - \gamma\delta.$$

Since from Eq.(16) it follows that $x_i = \lambda_{i0}/\sqrt{N}\sigma_1 - \gamma$, substituting this expression in Eq.(21) we obtain the sought expression for the marginal distribution of quasienergies. In the minimum the mean and the variance of this distribution are

$$\bar{\lambda}_i = \frac{1}{N}|E_0|, \qquad \sigma_\lambda^2 = N\sigma_1^2(1-\gamma\delta). \tag{22}$$

The dependence of the quantity $1-\gamma\delta$ on the reduced minimum depth $\gamma \approx |E_0|/N^{3/2}\sigma_T$ can be obtained solving equation (20) numerically. This dependence is shown in Fig.1. As we see, when

$\gamma$ increases the value of $1-\gamma\delta$ increases also. In the most interesting region $\gamma \sim 1.5$ we have $1-\gamma\delta \sim 0.55$. Many computer simulations performed on random matrices show that in the minimum we can estimate the energy as

$$|E_0| \approx 1.5 \cdot N^{3/2} \sigma_T, \qquad (23)$$

and for the energy of the ground state the estimate

$$|E_0| \approx 1.6 \cdot N^{3/2} \sigma_T$$

is true. In other words, in general we can use the value $\gamma \sim 1.55$ as an estimate of the parameter $\gamma$. Consequently, in the minimum the estimate for the variance of the quasienergy distribution is

$$\sigma_\lambda^2 \approx 0.55 \cdot N \sigma_T^2.$$

To verify the estimate for $\sigma_\lambda^2$ we performed a great number of computer simulations using random matrices of different types. They are matrices with the uniform or normal distributions of their elements, the Hebbian matrices and matrices of the Edwards-Anderson model.

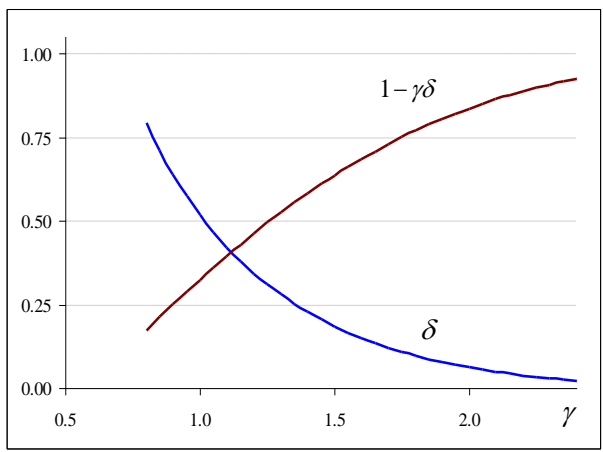 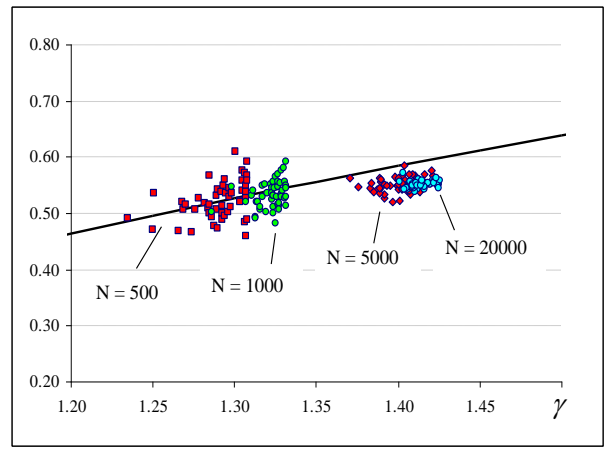

Fig.1. The dependence of the values of $\delta$ and $1-\gamma\delta$ on the normalized minimum depth $\gamma$, defined in Eq.(17).

Fig.2. The dependence of the variance $\overline{\sigma}_\lambda^2$ on the minimum depth $\gamma$. The full line corresponds to Eq.(23). Markers show experimental values for $N$=500, 1000, 5000, 20000.

In our computer simulations, we generated 200 matrices of the given type, and using the greedy Monte Carlo algorithm, we found $10^6$ minima. For each of the minima we defined $\sigma_\lambda^2$ and then averaged the obtained results over all the minima. In our computer simulations we use the matrices of dimensionalities $N \times N$, and $N$ varied from $N = 10^2$ to $N = 2 \cdot 10^4$. The results of our simulations are in good agreement with Eq.(22). In Fig.2 we show the results of simulations for the matrices of the Sherrington-Kirkpatrick model and dimensionalities

$N = 500, 1000, 5000, 20000$. The solid line corresponds the equation (22) for the normalized value $\bar{\sigma}_\lambda^2 = \sigma_\lambda^2 / N\sigma_T^2$. We used the same normalization when presenting the results of simulations.

Concluding this section, let us summarize the obtained results:

**i)** If the configuration corresponding to the energy minimum is chosen as $\mathbf{S}_0$, in Eqs (8) and (10) we have to substitute the estimates $\sigma_\lambda^2 \sim 0.6 \cdot N\sigma_T^2$ and $\bar{\sigma}_\lambda^2 \sim 0.6$ in the expression for the variance $\sigma_n^2$, respectively.

**ii)** With moving away from the minimum, the variance $\sigma_\lambda^2$ increases rapidly. Then if an arbitrary point is chosen as the initial configuration $\mathbf{S}_0$, when estimating the variance $\sigma_n^2$ in Eqs. (8) and (10) we have to use the values $\sigma_\lambda^2 = N\sigma_T^2$ and $\bar{\sigma}_\lambda^2 = 1$, respectively.

### 4. Computer simulations

Before describing our computer simulations, let us agree on definitions and notations we use. First, for simplicity we replace the long expression "the distribution of the energy states belonging to the $n$-vicinity" by a shorter one "the energy distribution in the $n$-vicinity". Second, up to the end of this section when denoting the mean value of the energy in the $n$-vicinity $\Omega_n$, we use the notation $E_n$ in place of the notation $\bar{E}_n$ of the previous sections.

#### 4.1. The experimental conditions

For a given $\mathbf{T}$ и $\mathbf{S}_0$ the energy $E(\mathbf{S}^{(k)})$ can be assumed as a random variable since in the sum (5) the factor $a_i^{(k)} a_j^{(k)}$ randomly takes the values $\pm 1$. The distribution of energies $E(\mathbf{S}^{(k)})$ depends on the type of the matrix $\mathbf{T}$, the initial configuration $\mathbf{S}_0$ and the number $n$ of the vicinity $\Omega_n$.

In the present paper, we restrict ourselves with matrices known in physics. The model traditionally used when checking new methods of calculation of the partition function is the Ising model on a hypercube lattice. The lattice dimension is denoted as $D$, and they distinguish one-dimensional ($D=1$), planar ($D=2$) and cubic ($D=3$) Ising models. Usually one takes into account the interaction between the nearest neighbors only. Thus, each spin interacts with $2D$ other spins and the connections between all these spins are the same. In the row of the $N$-dimensional matrix $\mathbf{T}$ there are only $2D$ equal nonzero elements. When $N \to \infty$ such a matrix is a substantially sparse one. The other analyzed matrix corresponds to the $D$-dimensional Edwards-Anderson model of the spin glass. This matrix resembles the $D$-dimensional Ising matrix, however the connections between spins are not the same, but they are random numbers generated with the aid of the standard normal distribution. When $N \to \infty$ this is the sparse matrix too. On the contrary, in the Sherrington-Kirkpatrick model, each spin is connected with all the other spins, and the connections are random numbers generated by the standard normal distribution. In this case, we deal with massive matrices (not sparse ones). We varied the dimensionality of the matrices in the limits $N \sim 10^2 - 10^4$.

In our computer simulations as an initial configuration $\mathbf{S}_0$ we chose either a random configuration, or the ground state. If it was not possible to determine the ground state, we used a deep energy minimum. In the Ising model, the ground state is the configuration $\mathbf{e} = (1,1,...,1)$. We mark out the initial configuration for this model, because if we set $\mathbf{S}_0 = \mathbf{e}$ the variance $\sigma_\lambda^2$ in Eq.10 is equal to zero: $\sigma_\lambda^2 = 0$. This simplifies the expressions (8) and (10) for the variance $\sigma_n^2$ significantly. Moreover, when $\mathbf{S}_0 = \mathbf{e}$, in each the vicinity $\Omega_n$ the minimal value of the energy asymptotically tends to $E_0$: $\min_{\mathbf{S} \in \Omega_n} E(\mathbf{S}) \to E_0 \; \forall n$. It is a very useful property since it is necessary to know the lower energy boundary $\min_{\mathbf{S} \in \Omega_n} E(\mathbf{S})$ when deriving the state equation (see [1,2]). Though the aforesaid relates to the Ising model on the hypercube, we assume that if we choose $\mathbf{S}_0$ as a ground state, for the matrices of other types we also can identify the lower boundary $\min_{\mathbf{S} \in \Omega_n} E(\mathbf{S})$ with $E_0$.

The relative quantity $x = n/N$ is the convenient characteristics for the distance between $\mathbf{S}_0$ and $\Omega$-vicinity. We performed all our simulations in the absence of the magnetic field ($\mathbf{H} = 0$), and consequently we varied the value of $x$ from 0 to ½.

Let us explain what we show in our figures. At first we set $(N \times N)$-matrix $\mathbf{T}$, the initial configuration $\mathbf{S}_0$ and the value of the parameter $x$. Next, we generated a large number $K \sim 10^6 - 10^7$ of random configurations from the $n$-vicinity ($n = xN$), calculated their energies $\{E_i\}_{i=1}^M$ and the frequencies of occurrence $\{K_i\}_{i=1}^M$: $\sum_{i=1}^M K_i = K$. Different values of the energies we denoted as $E_i$. The number of different energies $M$ is much less than the number of random configurations $K$: $M \ll K$. Since in the Ising model the energies formed a set of discrete values with a constant distance between the nearest energies, in our simulations we obtained not a histogram but a discrete distribution.

To compare the obtained discrete energies with the standard normal density of probabilities, we transformed the energies $E_i$ into $(E_i - E_n)/\sigma_n$, where $E_n$ and $\sigma_n^2$ were defined by Eqs.(8). The values of the standard normal density $p_i^{(th)}$ are:

$$p_i^{(th)} = \frac{1}{\sqrt{2\pi}} \exp\left[-\frac{1}{2}\left(\frac{E_i - E_n}{\sigma_n}\right)^2\right], \qquad i = 1,...,M.$$

The sum of all $p_i^{(th)}$ differs from 1: $\sum_{i=1}^{M} p_i^{(th)} = c_0 < 1$. Tough when comparing the theoretical probabilities $p_i^{(th)}$ with the experimental probabilities of occurrences of energies $E_i$ it is necessary to normalize them by the constant $c_0$:

$$p_i^{(ex)} = c_0 \frac{K_i}{K}, \quad i = 1, ..., M. \tag{24}$$

In all the figures of this section on the abscissa axes we plot the transformed energies $(E_i - E_n)/\sigma_n$. It is necessary to have in mind that the integers 1, 2, 3, 4... on these axes correspond to the original energies $E_i$ that are spaced at $\sigma_n$, $2\sigma_n$, $3\sigma_n$, $4\sigma_n$... from the mean value of the energy $E_n$. This remark will be useful when discussing the results for the Ising models on the lattices of different dimensionalities $D$ (see Section 5).

Each figure consists of a number of panels each of which has the upper and lower parts. In each upper part, markers indicate the experimental probabilities (24), and the solid line is the graph of the standard normal density. When all the markers are rather close to the solid line, our experimental probabilities are in good agreement with the normal distribution. In each lower part of a panel, we show another graph characterizing the distinctions of the experimental probabilities from the normal density. Here markers show the dependence of the common logarithm $\lg(p_i^{(ex)}/p_i^{(th)})$ on the energy $(E_i - E_n)/\sigma_n$. When $p_i^{(ex)} \approx p_i^{(th)}$, the logarithm approximately equals to zero. However, when $p_i^{(ex)}$ and $p_i^{(th)}$ differ noticeably the value of the logarithm shows to what degree these probabilities differ. In what follows we show that when the energies belong to the interval $|E_i - E_n|/\sigma_n \leq 3$, as a rule the values of the logarithm are close to zero. Outside this interval they differs from zero noticeably.

**4.2. Results for the Ising model.**

4.2.1. In Fig.3. we show the results of four simulations for the $3D$ Ising model. The dimesionality of the matrix was $N = 10 \times 10 \times 10 = 1000$. We used the ground state ($\mathbf{S}_0 = \mathbf{e}$) as the initial configuration, the value of the parameter $x$ changed from 0.5 to 0.005. The number of random configurations used in each simulation was $K = 10^7$. The four panels show how with decrease of $x$ the distribution of the energy increasingly differs from the normal one.

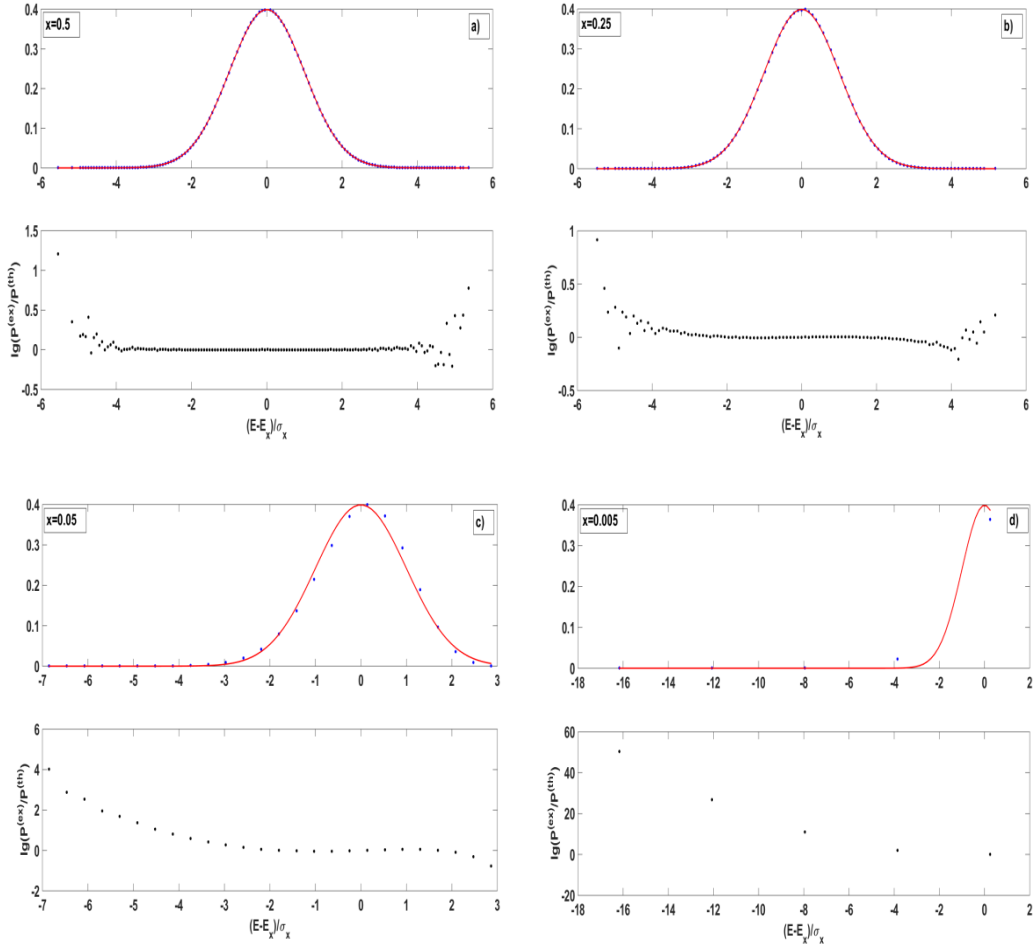

**Fig.3.** 3D Ising model, $N = 10 \times 10 \times 10 = 1000$, the ground state is the initial configuration: $\mathbf{S}_0 = \mathbf{e}$, $K = 10^7$. For different values of the parameter $x$ it is shown how the experimental probabilities $p_i^{(ex)}$ correspond to the standard Gaussian curve (see Sec.4.2.1): a) $x = 0.5$, b) $x = 0.25$, c) $x = 0.05$, d) $x = 0.005$.

The panel (a) shows that when $x = 0.5$ the normal density describes the experimental probabilities perfectly well. According to the lower part of this panel inside the interval $[-4, 4]$, the coincidence of the experimental results with the normal density is very good. A discrepancy takes place only outside this interval, but it is not very large. The panel (b) shows that even when $x = 0.25$ the experimental points agree with the normal density nearly as good as for larger values of $x$. The panel (c) corresponds to the case of a small value of the parameter $x$: $x = 0.05$. Here we clearly see systematic discrepancies between the experimental probabilities and the normal density. The experimental distribution has a long left tail where the experimental probabilities exceed the theoretical values by 1-2 orders of magnitude. Note, here the number of different energies $M$ is much less than on the panels (a) and (b). Finally, on the panel (d) we present the graphs for even smaller value of the parameter $x$: $x = 0.005$. In this case, we have five different values of the energy only. When the energy is close to zero, the discrepancy between the theory and experiment is not very large, but on the left tail the experimental probabilities exceed the theoretical values by many orders of magnitude (see the lower part of the panel).

The main conclusion following from these experiments states that the distribution of energies is not a normal one in the vicinities $\Omega_n$, corresponding to small values of the parameter $x = n/N$. For $N = 1000$ it takes place when $x \leq x_c$, where the critical value $x_c \approx 0.15$.

4.2.2. When increasing the dimensionality $N$ the critical region where the normal density describes the experimental data badly also decreases and it tends to zero when $N \to \infty$. For the Ising model the proper estimates can be easily done. Indeed, we can use the normal density to approximate the experimental probabilities when the peak of distribution is far enough from both the upper and lower boundaries of the distribution. Only in this case the normal curve is not cut off both from the left and from the right. For the Ising matrices the maximal and minimal values of the energy in the $n$-vicinity of the ground state has the form:

$$E_{min} \approx E_0 + 4q/\sqrt{n}, \quad E_{max} = E_0 + 4qn, \quad E_0 = -Nq, \quad E_n = E_0 \cdot (1-2x)^2,$$

where $q = 2D$ is the number of the nearest neighbors. The peak of distribution is distant from the boundaries when

$$E_n - E_{min} \gg \sigma_n, \quad E_{max} - E_n \gg \sigma_n. \tag{25}$$

From figures 3-5 we see that the energy distribution is cut off from the right, where the value of $x = n/N$ is small. Then defining the critical value $x_c$ has to be defined from the second inequality in Eq.(25) we obtain

$$x \gg x_c = \frac{1}{\sqrt{qN}}.$$

Thus, when $x < x_c$ it is not correct to approximate the real distribution by the normal curve. However, in the asymptotic limit $N \to \infty$ the size of this region tends to zero.

Graphs in Fig.4 that show the results of simulations for $1D$ Ising model and $K = 10^6$, demonstrate this statement clearly. As before, we take $\mathbf{S}_0 = \mathbf{e}$, however now we fix the value of the parameter $x$ ($x = 0.05$) and vary the dimensionality of the problem from $N = 500$ up to $N = 20000$. We do not analyze in details all the graphs shown in four panels of this figure. We note only that the larger the dimensionality $N$, the better experimental points fit the normal curve, and when $N = 20000$ the correspondence between the experimental probabilities and the normal density is rather good. We can be sure that for larger dimensionality ($N \sim 10^5$) the experimental points fit the normal curve almost perfectly.

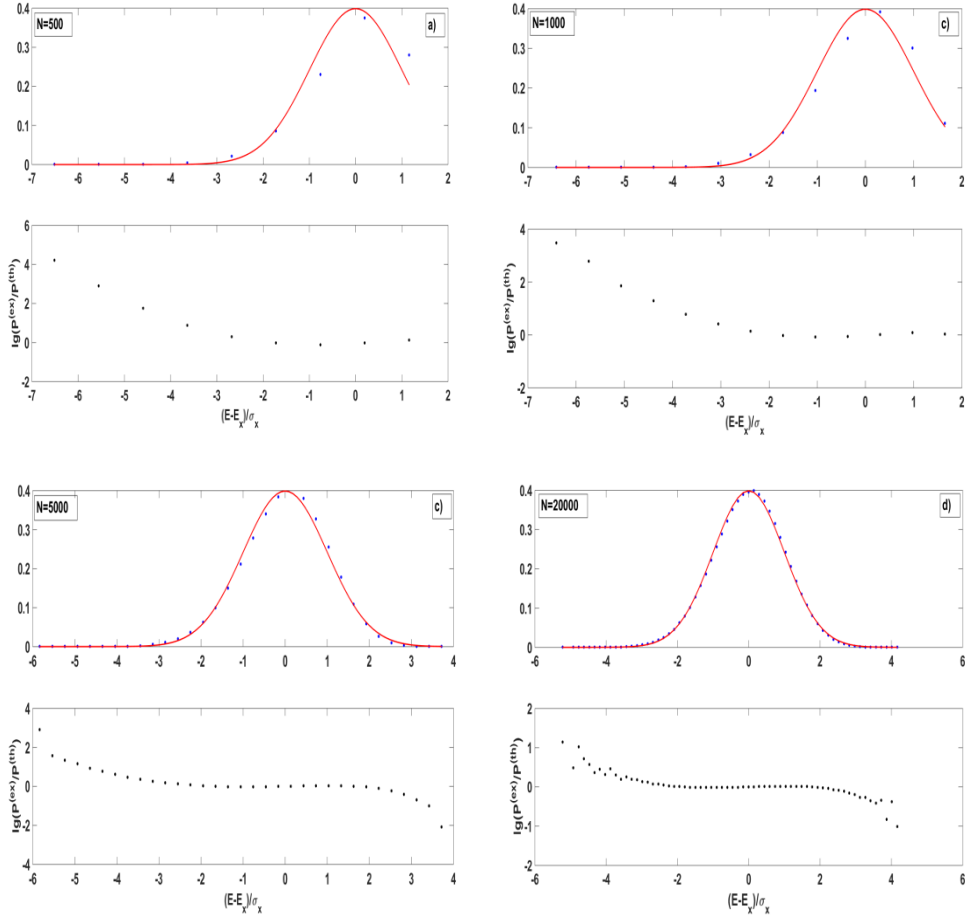

**Fig.4**. 1D Ising model, the ground state is the initial configuration: $\mathbf{S}_0 = \mathbf{e}$, $x = 0.05$, $K = 10^6$. For different dimensionalities $N$ it is shown how the experimental probabilities $p_i^{(ex)}$ correspond to the standard Gaussian curve (see Sec.4.2.2): a) $N = 500$, b) $N = 1000$, c) $N = 5000$, d) $N = 20000$.

Summarizing this subsection, we can state that in the limit $N \to \infty$ the size of the critical region where the normal approximation is not true tends to zero: $x_c \to 0$.

4.2.3. In Fig.5 we show the results of simulations for the 3D Ising model for the matrix of very large dimensionality $N = 33 \times 33 \times 33 = 35937$. We took the ground state as the initial configuration ($\mathbf{S}_0 = \mathbf{e}$) and varied the parameter $x$ from 0.003 to 0.139. In different simulations the number of the tests $K$ varied from $3 \cdot 10^6$ to $4 \cdot 10^7$. We see that when the dimensionality is so large, the experimental probabilities correspond to the normal curve rather good beginning from $x = 0.028$. When the values of $x$ are even larger the agreement between the theory and simulations can be regarded as nearly perfect (see two lower panels of this figure). Again, we see that when $N$ increases the size of the critical region where the normal approximation for the energy distribution is not correct, decreases.

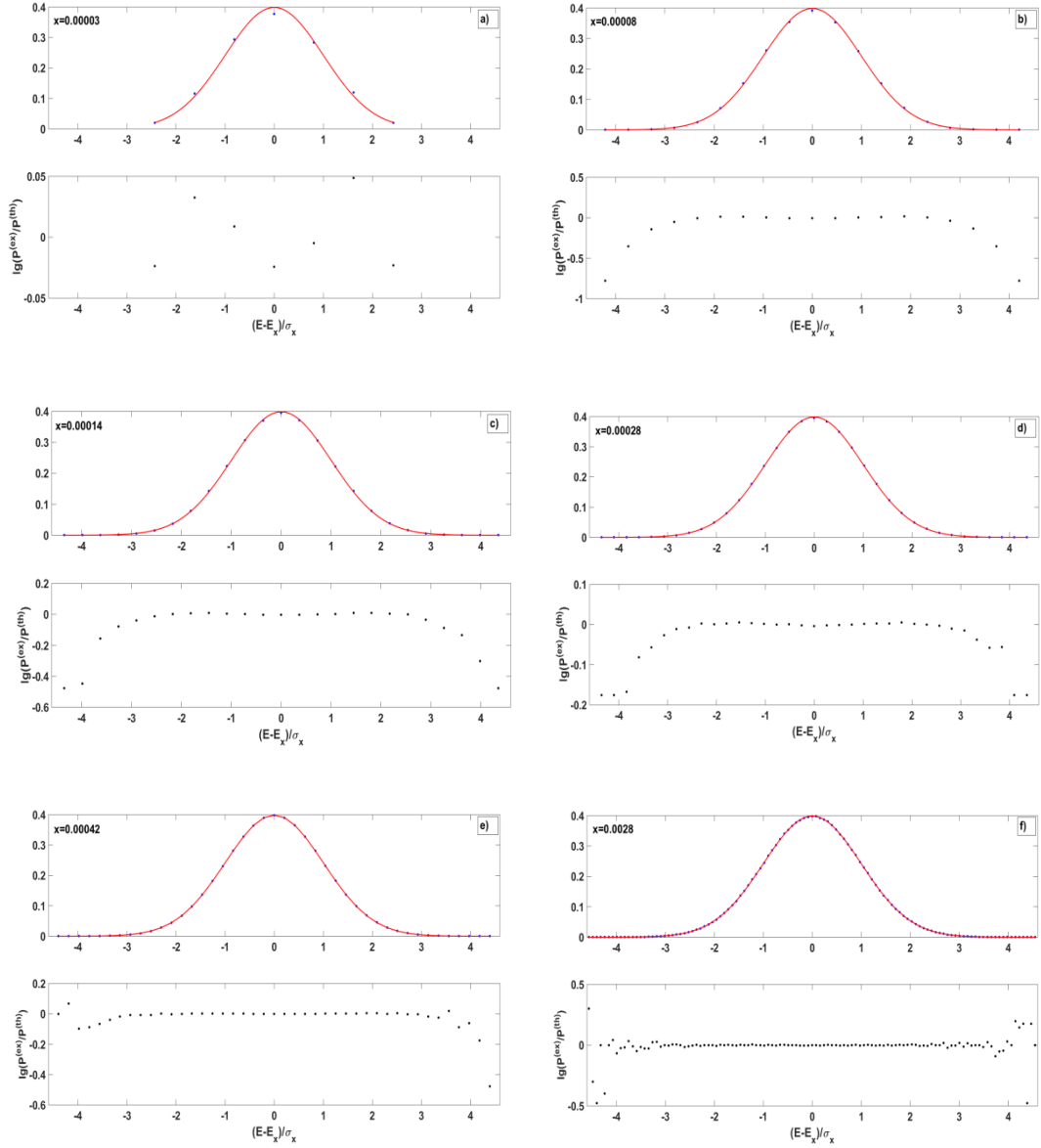

**Fig.5.** $3D$ Ising model, $N = 33 \times 33 \times 33 = 35937$, the ground state is the initial configuration: $\mathbf{S}_0 = \mathbf{e}$, $K \sim 5 \cdot 10^6$. For different values of the parameter $x$ it is shown how the experimental probabilities $p_i^{(ex)}$ correspond to the standard Gaussian curve (see Sec.4.2.2): a) $x = 0.003$, b) $x = 0.006$, c) $x = 0.008$, d) $x = 0.014$, e) $x = 0.028$, f) $x = 0.139$.

4.2.4. In Fig.6 we show the experimental results for the $3D$ Ising model with the matrix of the same large dimensionality $N = 35937$. However now a random configuration plays the role of the initial one: $\mathbf{S}_0 = \mathbf{S}_{\text{rand}}$. The value of $x$ changes from 0.00003 to 0.0028, in different simulations the numbers of tests $K$ vary from 3 to 9 millions. The panel (a) corresponds to the 1-vicinity $\Omega_1$ of the initial configuration ($x = 0.00003$); the panel (b) corresponds to the 3-vicinity $\Omega_3$ ($x = 0.00008$) and so on, the panel (f) corresponds to the vicinity $\Omega_{100}$ ($x = 0.0028$). The graph in the panel (b) shows that even for the 3-vicinity of the initial configuration the experimental probabilities fit the normal curve in the rather wide interval of energies. Beginning from $x = 0.00028$ the agreement between the experimental probabilities and the normal density is almost perfect for the energies from the interval $|E_i - E_n|/\sigma_n \leq 3$.

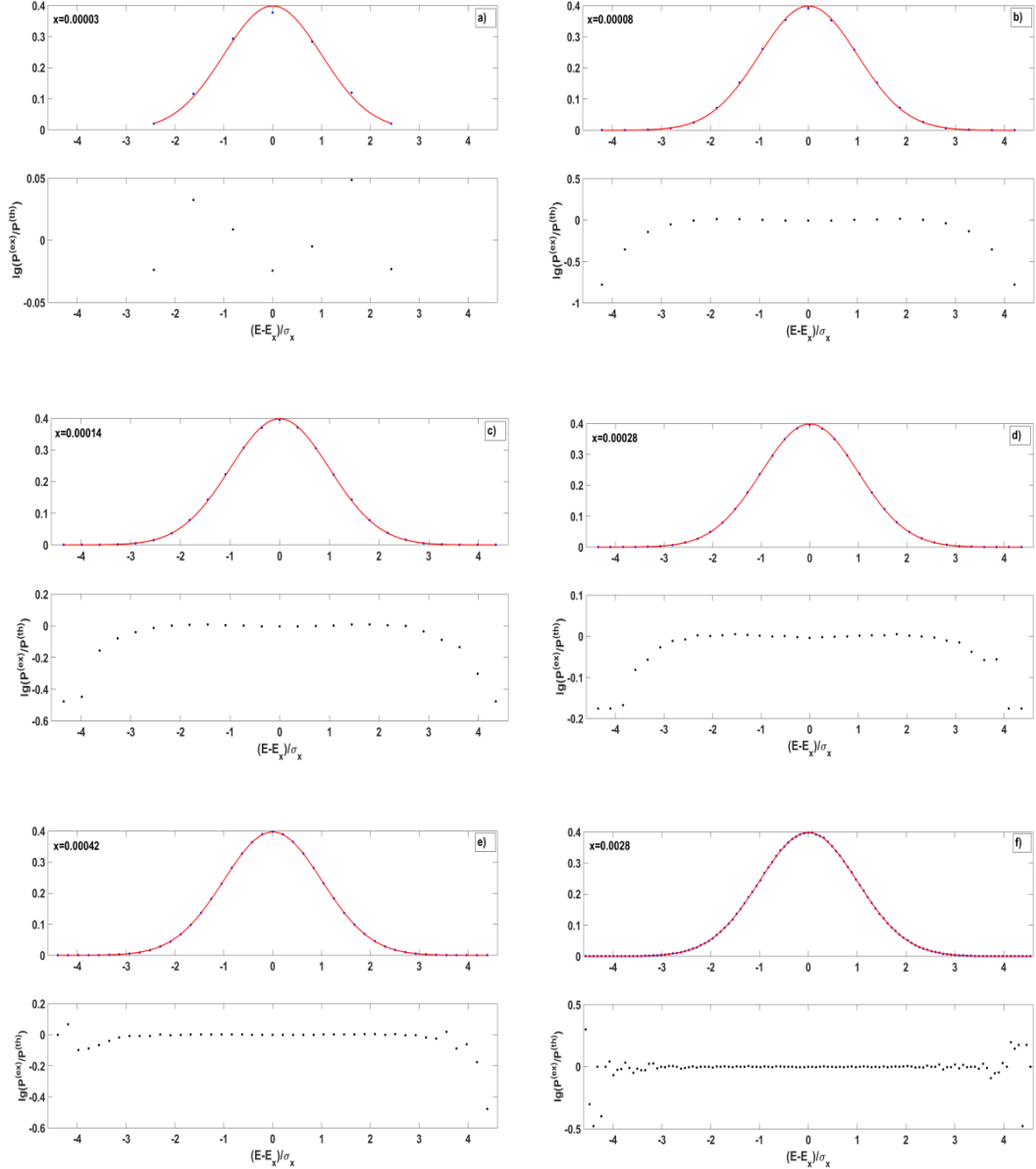

**Fig.6.** 3D Ising model, $N = 35937$, $\mathbf{S}_0$ is a random configuration, $K \sim 5 \cdot 10^6$. For different values of the parameter $x$ it is shown how the experimental probabilities $p_i^{(ex)}$ correspond to the standard Gaussian curve (see Sec.4.2.2): a) $x = 0.00003$, b) $x = 0.00008$, c) $x = 0.00014$, d) $x = 0.00028$, e) $x = 0.00042$, f) $x = 0.0028$.

We see that in the case of the Ising model, the choice of a random configuration as $\mathbf{S}_0$ significantly extends the region, where our basic assumption about the normal type distribution of the energies in the $n$-vicinity $\Omega_n$ is true. It is easy to understand that this is especially true for matrices whose elements are random numbers. However, with this choice of the initial configuration it is difficult to estimate the lower boundary of the energies for the $n$-vicinity, and this makes difficulties when deriving the state equation [1], [2].

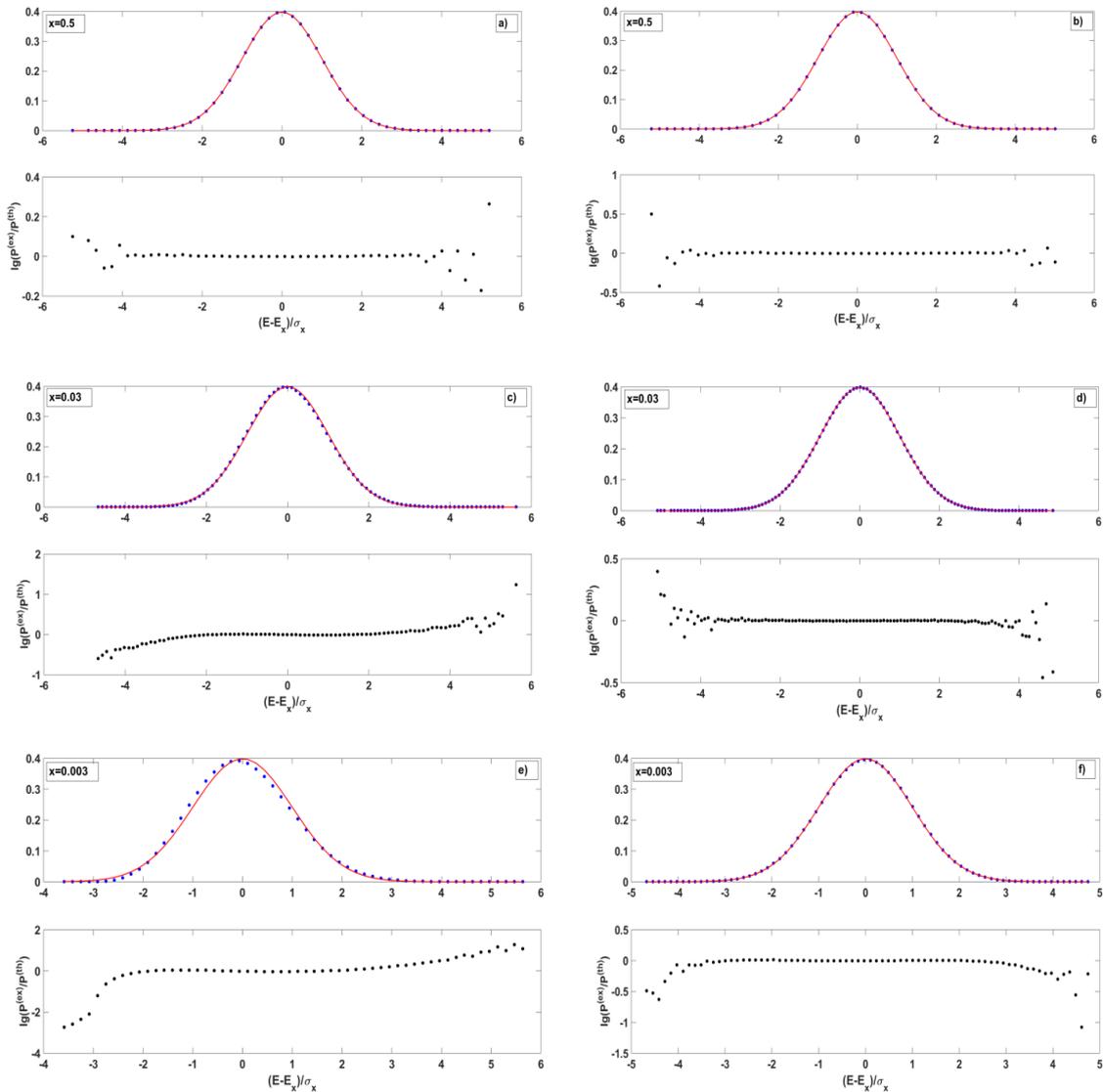

**Fig.7.** The Edwards-Anderson model $D = 3$, $N = 1000$, $K = 2 \cdot 10^7$, $x = 0.5, 0.03$ and 0.003 (downward). In the left panels the ground state is the initial configuration $\mathbf{S}_0$, in the right panels $\mathbf{S}_0$ is a random configuration. All the rest the same as in Figs.3-6.

### 4.3. Results for the spin glass models.

4.3.1. In Fig.7 we show the results of computer simulations for the three-dimensional Edwards-Anderson model of the spin glass. The dimensionality of the matrix was $N = 10 \times 10 \times 10 = 1000$, its nonzero elements were random numbers generated with the aid of the standard normal distribution. The parameter $x$ was $x = 0.5, 0.03$ and 0.003. On the left graphs we took the ground state as the initial configuration $\mathbf{S}_0$, and the right graphs correspond to a random configuration used as $\mathbf{S}_0$. The number of tests was the same for all the simulations: $K = 2 \cdot 10^7$.

When $x = 0.5$, regardless of configuration used as the initial configuration $\mathbf{S}_0$ both graphs demonstrate good agreement of the experimental distribution with the normal density, see panels (a) and (b) of the figure. Then when the value of the parameter $x$ becomes small ($x = 0.03$) and very small ($x = 0.003$) the graphs behave themselves in different ways depending on the initial configuration $\mathbf{S}_0$. If we take a random configuration as $\mathbf{S}_0$ (see graphs on the panels (d) and (f)), as before the experimental probabilities fit the normal distribution rather good. However, if the ground state is taken as $\mathbf{S}_0$ (see graphs on the panels (c) and (e)), the decrease of $x$ is accompanied by a growing deviation of the experimental values from the normal curve. The distribution becomes asymmetric and its left tail is cut off. The global minimum serves as the lower boundary for the energies of configurations close to $\mathbf{S}_0$.

4.3.2. In Fig.8 we present the results for the Sherrington-Kirkpatrick model of the spin glass. The dimensionality of the matrix was $N = 1000$, and $N(N-1)/2$ of its matrix elements were random numbers generated with the aid of the standard normal distribution. The parameter $x$ took the values $x = $ 0.5, 0.01 and 0.003. As in the previous figure the left graphs correspond to the ground state used as the initial configuration $\mathbf{S}_0$, and the right ones to the case of a random configuration used a $\mathbf{S}_0$. In all the simulations the numbers of tests were $K = 2 \cdot 10^7$.

In general, the results of simulations is much alike the results obtained for the Edwards-Anderson model (see Fig.7). When we take a random configuration as $\mathbf{S}_0$, the experimental data are in good agreement with the normal density even for very small values of the parameter $x = 0.003$. On the contrary, when we take the ground state as the initial configuration $\mathbf{S}_0$, even for $x = 0.01$ the difference between the experimental probabilities and the normal density distribution becomes noticeable (see the lower curve in the panel (c)). When the parameter $x = 0.003$, the difference increases (see graphs on the panel (e)) of the figure). We emphasize that when increasing the dimensionality of the matrix $N$ substantively, we can achieve good agreement between the experiment and the normal density for small values of the parameter $x$, at least for the energies from the interval $|E_i - E_n|/\sigma_n \leq 3$.

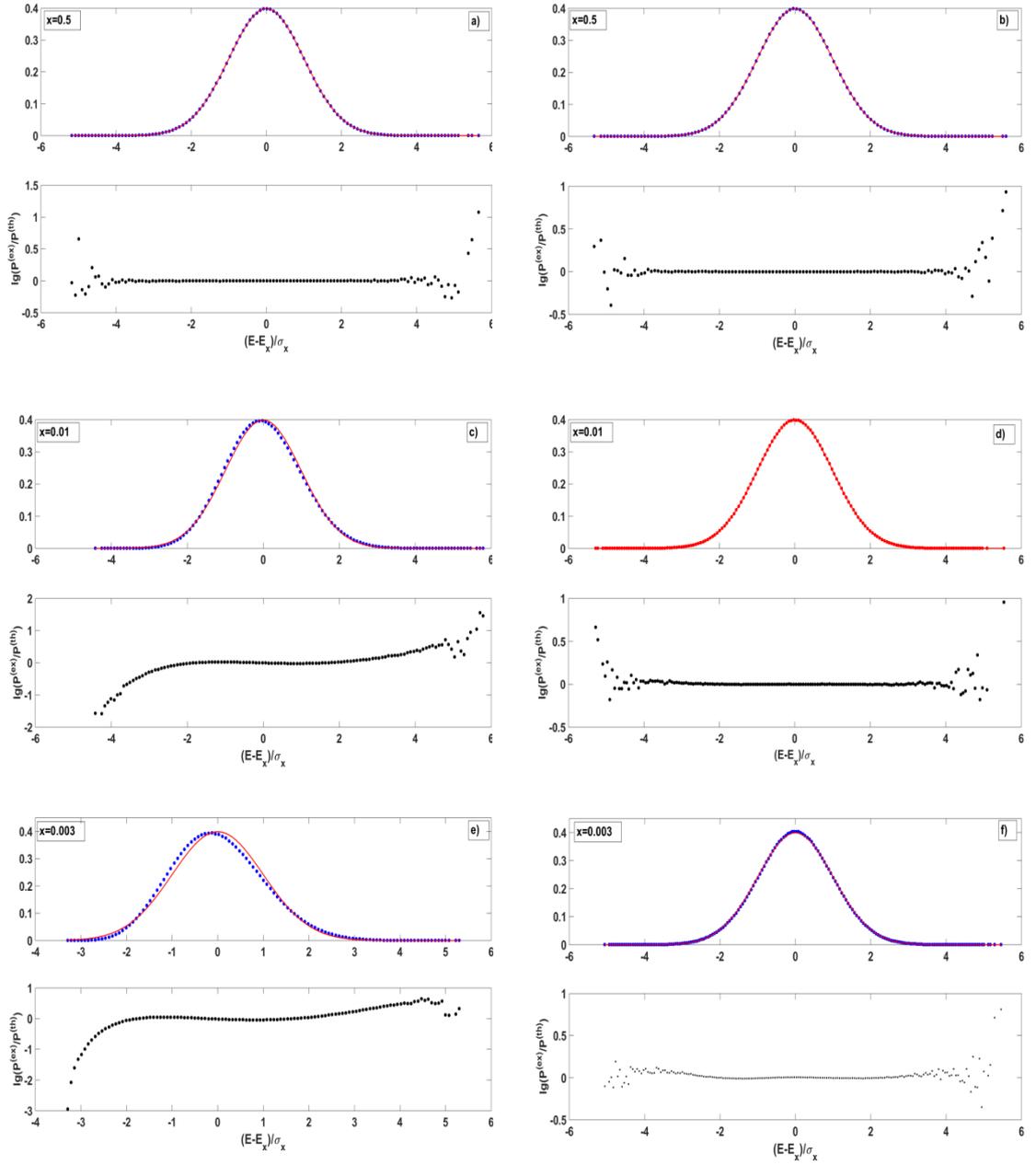

**Fig.8.** The Sherrington-Kirkpatrick model, $N = 1000$, $K = 2 \cdot 10^7$, $x = 0.5$, 0.01 and 0.003 (downward). In the left panels the ground state is the initial configuration $\mathbf{S}_0$, in the right panels $\mathbf{S}_0$ is a random configuration. All the rest the same as in Fig.7.

## 5. Discussion and prospects

In our approach, the most important is the assumption that the energy distribution in the $n$-vicinity $\Omega_n$ can be approximated accurately by the normal distribution. Qualitatively the justification of this supposition is as follows. We normalize the energy in such a way that it becomes independent of the dimensionality $N$. For example, in the case of the Ising model it is sufficient to pass to $\tilde{E} = E/N$, i.e. to the energy per spin, and for systems with random matrices we use the value $\tilde{E} = E / \sqrt{N \cdot \mathrm{Sp}\, \mathbf{T}^2}$. Then using Eqs.(10)-(11) we can show that when $N$ increases, the variance of the variable $\tilde{E}$ tends to zero as

$$\tilde{\sigma}_n^2 = \frac{c(m)}{N},$$

where $c(m)$ is a function of magnetization $m$ that is independent of the dimensionality $N$. In the case of the Ising model it is

$$c(m) = 2q(1-m^2)^2,$$

and for the models with random matrices

$$c(m) = 2 \cdot (1-m^2)\left[(1-m^2) + 2m^2 \bar{\sigma}_\lambda^2\right].$$

For the given normalization in the case of the Ising model the value of $\tilde{E}$ changes inside the finite interval $[-q, +q]$ and for the models with random matrices the interval is $[-1.6, +1.6]$. When $N \to \infty$ the variance of $\tilde{E}$ tends to zero as $1/N$. This means that in this limit the distribution of $\tilde{E}$ turn into the delta-function that is approximated perfectly by the normal distribution when its variance tends to zero. The expression (A5) for the third moment of the energy distribution confirms the aforesaid statement: for our normalization in the limit $N \to \infty$ the term $R$ in Eq.(A5) is vanishingly small.

We emphasize that the approximation of the energy distribution by the normal one is true only for its central part in a vicinity of the mean value $E_n$. In the general case, we cannot say anything definite about its tails. For examined models, our simulation showed the following.

**i)** If we take the ground state as the initial configuration $\mathbf{S}_0$, the energy distributions are asymmetric ones with respect to the mean value $E_n$:

- In the case of the Ising models, the distributions have a thick left tails directing to the global minimum and the right ones quickly falling off in the direction of large energies. The asymmetries of distributions are the most visible for small values of the parameter $x = n/N$ and it decreases when the number of the nearest neighbors $q$ increases.

- In the case of models with random matrices, the reverse is true: the distributions have quickly falling off left tails and thick right tails. The asymmetries of distributions are the most visible for small values of $x = n/N$.

**ii)** If we take a random configuration as the initial configuration $\mathbf{S}_0$, the distributions are symmetric with respect to $E_n$:

- In the case of the Ising models, the distributions have thin quickly falling off tails (the left and the right ones). The less the parameter $x = n/N$, the more the tails differ from of the normal distribution.

- In the case of the models with random matrices, the reverse is true: the distributions have thick tails. The difference of these tails from the normal distribution increases when the parameter $x = n/N$ increases.

"Thick" and "thin" tails mean that our experimental curves lie above or under the theoretical (the normal) curve, respectively.

Finally, let us discuss how this approach can be used in the problem of the partition function calculation. When replacing in the summation of the exponent $\exp[-\beta E(\mathbf{S})]$ over the $n$-vicinity by integration of $\exp(-\beta E)$ times the normal density, we can obtain the state equation with the aid of the standard calculating methods. Solution and examination of this equation provides us with such characteristics as the critical temperature, the kind of the phase transition, the values of the critical indices and so on. In [1, 2] we have done this for the $D$-dimensional Ising models and the obtained results were ambiguous. When $D \geq 3$, the calculated critical temperatures were in good agreement with generally accepted values, but for $D < 3$ our results differed substantially from the well-known exact results (see Table.1.)

Table 1. Critical values of the reciprocal temperature $\beta_c$.

|  | $D = 2$ | $D = 3$ | $D = 4$ |
|---|---|---|---|
| Exact value[1] | 0.4407 | 0.2216 | 0.1489 |
| Our model | 0.3912 | 0.2146 | 0.1464 |

We see that if $D \geq 3$ the obtained critical values are close to the standard results. Moreover, when we increase $D$ indefinitely that is when increasing the number of nonzero matrix elements in the row, in the limiting case $q = N - 1$ we in fact return to the classical mean field model. This well-known model is described by the Bragg-Williams equation [23]. In the limit $q = N \to \infty$ we obtain the same equation. In other words, the more the number of the nonzero matrix elements q, the more accurate is our method of the partition function calculation.

On the contrary, for the two-dimensional Ising model our value of the critical temperature differs from the exact one by over 10% (see Table.1). It is even worth that when using our approach a jump of magnetization (the phase transition of the first kind) takes place prior the phase transition of the second kind. This is not agree with the generally accepted phase diagram for the two-dimensional Ising model. This means that in this case our method provides the structurally wrong solution. We do not present the results for the one-dimensional Ising model in Table.1. However, they are also "structurally wrong". From our equations, it follows that the phase transition of the first kind takes place whereas it is well known, that there is no phase transition in the one-dimensional system. We hope that we understand the reason of these faults.

---
[1] For $D = 3$ and $D = 4$ as the exact values we took the results of computer simulations from [21] and [22], respectively.

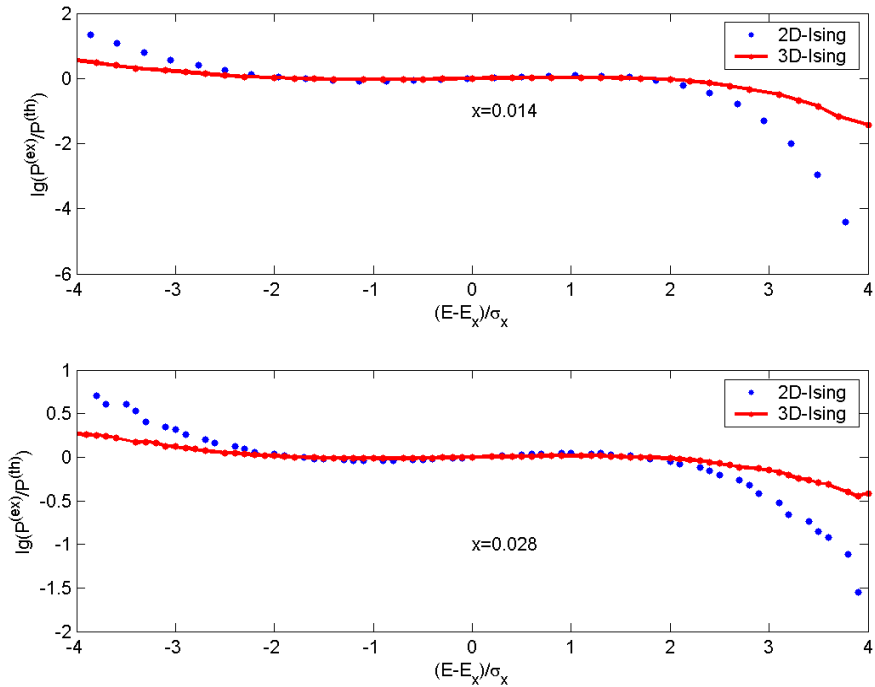

**Fig.9.** Comparison of the values of $\lg\left(p_i^{(ex)}/p_i^{(th)}\right)$ for $2D$- and $3D$-Ising matrices for the same dimensionality ($N = 36000$) and the same values of the parameter $x$.

To begin with, the results of the computer simulations allows us to define our main supposition more accurately: for each finite value of $x$ we can find such a value of $N$ that the energy distribution in the $n$-vicinity $\Omega_n$ inside the interval $E_n \pm k\sigma_n$ ($n = xN$) can be described reasonable well by the normal distribution. The values of the factor $k$ obtained in our computer simulations varied nearby $k \approx 2-4$. Note, for the Ising model $\sigma_n \sim \sqrt{q}$ and, consequently, the more the dimensionality of the hypercube $D = q/2$, the wider the interval of energies where the energy distribution can be assumed as the normal one (the value of the parameter $x$ is fixed). Then, the more $q$, the wider the interval of energies where our approximation is correct, and the more accurate is our approach.

In Fig.9 we compare the values of $\lg\left(p_i^{(ex)}/p_i^{(th)}\right)$ for $2D$- and $3D$- Ising matrices for the same dimensionality ($N = 36000$) and the same values of the parameter $x$. In the upper panel, we present graphs for $x = 0.014$ and in the lower one for $x = 0.028$. In both panels, the curves corresponding to the $3D$- Ising model are flatter than the curves corresponding to the $2D$-Ising model. The difference in the behaviors of the curves becomes evident beginning from $|(E - E_x)/\sigma_x| = 2$. At the tails this difference is rather large. This means that in the case of the $3D$-Ising model the normal distribution approximates the energy distribution in the $n$-vicinity far better than in the case of the $2D$-Ising model.

Of course, it would be good to understand why the critical value of $D$ is equal to three and why our approach does not work for the lattices of less dimensionalities. We will try to answer this question in our following studies. In addition, it is important to understand how to describe the tails of the energy distributions in the $n$-vicinity. Though we know that when $n = xN$ and $x \neq 0$, the normal distribution is good for describing the central part of the energy distribution, it is unclear how to characterize its tails. Most likely that the asymmetry of the energy distribution becomes increasingly important when $x$ decreasing from ½ toward zero. The behavior of the curves on our graphs indicate such a possibility. This question requires further analysis.

**Acknowledgments**

The work is supported by RBRF grants №15-07-04861 and №13-01-00540. The authors are grateful to Ya.M.Karandashev, V.M.Kryzhanovskiy and M.Yu.Malsagov who helped to design algorithms for calculation the energy spectrums as well as in computer simulations.

**Appendix**

Let $n$ components of a vector $\mathbf{A}^{(k)} = (a_1^{(k)}, a_2^{(k)}, ..., a_N^{(k)})$ are equal to -1, and the other ones are equal to +1. The vicinity $\Omega_n$ of the initial configuration $\mathbf{S}_0$ is defined by a set of all the vectors $\{\mathbf{A}^{(k)}\}$, when $k = 1, 2, ..., L$, $L = C_N^n$. We obtain components of a vector $\mathbf{S}^{(k)} \in \Omega_n$ multiplying the components of the initial configuration $\mathbf{S}_0$ and one of the vectors $\mathbf{A}^{(k)}$: $s_i^{(k)} = s_i^{(0)} a_i^{(k)}$. In Section 2, we note that to calculate the moments of the energy distribution we have to know how to average the quantities $a_{i_1}^{(k)} a_{i_2}^{(k)} ... a_{i_r}^{(k)} \chi_{i_1 i_2 ... i_r}$, where $\chi_{i_1 i_2 ... i_r} = 1$ when all the indices are distinct, and otherwise $\chi_{i_1 i_2 ... i_r} = 0$. In other words, the problem is to calculate

$$K_r = \overline{a_{i_1}^{(k)} a_{i_2}^{(k)} ... a_{i_r}^{(k)} \chi_{i_1 i_2 ... i_r}} \equiv \frac{1}{L} \sum_{k=1}^{L} a_{i_1}^{(k)} a_{i_2}^{(k)} ... a_{i_r}^{(k)} \chi_{i_1 i_2 ... i_r}$$

To determine the first three moments we need the values of $K_r$ with $r = 1,...,6$. We start with the simplest case $r = 1$. It is very simple to calculate $K_1 = \overline{a_i^{(k)}}$: for $n$ of $N$ cases the value $a_i^{(k)} = -1$, and $a_i^{(k)} = +1$ in other $N - n$ cases. Consequently, the mean value $\overline{a_i^{(k)}} = (N - 2n)/N$.

We proceed to calculation of $K_2 = \overline{a_i^{(k)} a_j^{(k)} \chi_{ij}}$. It is easy to understand that if $i \neq j$ the signs of $a_i^{(k)}$ and $a_j^{(k)}$ coincide in $n(n-1) + (N-n)(N-n-1)$ cases and the signs are opposite in other $2n(N-n)$ cases. The number of possible cases is equal to $N(N-1)$ and consequently, the mean value we are looking for is

$$K_2 = \overline{a_i^{(k)} a_j^{(k)} \chi_{ij}} = \frac{[n(n-1) + (N-n)(N-n-1)] - 2n(n-1)}{N(N-1)}.$$

The same arguments lead to the expressions for all the values $K_r$ we need:

$$K_1 = \frac{N - 2n}{N},$$

$$K_2 = \frac{(N - 2n)^2 - N}{N(N - 1)},$$

$$K_3 = \frac{(N - 2n)^3 - (N - 2n)(3N - 2)}{N(N - 1)(N - 2)}, \qquad (A1)$$

$$K_4 = \frac{(N - 2n)^4 - 2(N - 2n)^2(3N - 4) + 3N(N - 2)}{N(N - 1)(N - 2)(N - 3)},$$

$$K_6 = \frac{(N - 2n)^6 - 5(N - 2n)^4(3N - 17) + 180n(N - n)(N - 2n)^2 + O(N^4)}{N(N - 1)(N - 2)(N - 3)(N - 4)(N - 5)}.$$

We start the calculation of the moments of the energy distribution with the case of the absence of the magnetic field: $\mathbf{H} = 0$. In Section 2 for the first moment we obtained the relation $\overline{E}_n = K_2 \cdot E_0$. The first of two expressions (8) follows from this relation with account of Eq.(A1). Now let us calculate the mean value of the square of energy:

$$\overline{E_n^2} = \frac{1}{L} \sum_{k=1}^{L} E^2(\mathbf{S}^{(k)}).$$

The vector $\mathbf{S}^{(k)}$ we present in the form of the componentwise product of the vectors $\mathbf{A}^{(k)}$ and $\mathbf{S}^{(0)}$ and write the squared energy as

$$E^2(\mathbf{S}^{(k)}) = -\sum_{i=1}^{N} \sum_{j=1}^{N} \sum_{l=1}^{N} \sum_{r=1}^{N} T_{ij} T_{lr} s_i^{(0)} s_j^{(0)} s_l^{(0)} s_r^{(0)} \cdot a_i^{(k)} a_j^{(k)} a_l^{(k)} a_r^{(k)} \chi_{ij} \chi_{lr} \qquad (A2)$$

Since we know how to average the product of the components of the vector $\mathbf{A}^{(k)}$ with different indices, we add to Eq.(A2) the multiply identically equal to one and write the obtained expression as

$$\chi_{ij}\chi_{kr} = \chi_{ij}\chi_{kr}(\delta_{ik}+\chi_{ik})(\delta_{ir}+\chi_{ir})(\delta_{jk}+\chi_{jk})(\delta_{jr}+\chi_{jr}) = \\ = (\delta_{ik}\delta_{jr}+\delta_{ir}\delta_{jk})\chi_{ij}+(\delta_{ik}\chi_{ijr}+\delta_{ir}\chi_{ijk}+\delta_{jk}\chi_{ijr}+\delta_{jr}\chi_{ijk})+\chi_{ijkr}. \quad (A3)$$

The first term in the right-hand side of Eq.(A2) provides the contribution independent of the components $s_i^{(0)}$, the second term contains the product of two components $s_i^{(0)}$ and $s_j^{(0)}$, the third one includes the product of the four components. Substituting the expression (A3) in Eq.(A2) and summarizing over the all class $\Omega_n$, we obtain

$$\overline{E_n^2} = 2\mathrm{Sp}\,\mathbf{T}^2 + 4K_2\sum_{i=1}^{N}\sum_{j=1}^{N}(T^2)_{ij}s_i^{(0)}s_j^{(0)}\chi_{ij} + K_4\sum_{i=1}^{N}\sum_{j=1}^{N}\sum_{k=1}^{N}\sum_{r=1}^{N}T_{ij}T_{kr}s_i^{(0)}s_j^{(0)}s_k^{(0)}s_r^{(0)}\chi_{ijkr}, \quad (A4)$$

where $\mathrm{Sp}\,\mathbf{T}^2$ denotes trace of the matrix $\mathbf{T}^2$: $\mathrm{Sp}\,\mathbf{T}^2 = \sum_{i,j=1}^{N}T_{ij}^2$.

After rather simple calculations, rearranging the terms in Eq.(A4) we obtain more compact expression for $\overline{E_n^2}$. In the same way it is possible to obtain the expression for the third moment $\overline{E_n^3}$. Omitting intermediate calculations, we present the expressions for the three moments:

$$\overline{E}_n = E_0\frac{(N-2n)^2 - N}{N(N-1)},$$

$$\overline{E_n^2} = \overline{E}_n^2 + \sigma_n^2, \quad (A5)$$

$$\overline{E_n^3} = \overline{E}_n^3 + 3\overline{E}_n\sigma_n^2 + R,$$

where the variance $\sigma_n^2$ is given by the expression:

$$\sigma_n^2 = 2(1+K_4-2K_2)\sum_i\sum_j T_{ij}^2 + 4(K_2-K_4)\sum_i\lambda_i^2 + (K_4-K_2^2)E_0^2. \quad (A6)$$

We introduce the notations:

$$E_{30} = -\sum_{i,j}(T^3)_{ij}s_i^{(0)}s_j^{(0)}, \quad \lambda_i = \sum_j s_i^{(0)}T_{ij}s_j^{(0)},$$

$$R = 8(1-3K_2+3K_4-K_6)\cdot\mathrm{Sp}\,\mathbf{T}^3 + \\ + 2(K_2-2K_4+K_6)\left[8\sum_{i,j}T_{ij}^3 s_i^{(0)}s_j^{(0)} - 24\sum_i(T^2)_{ii}\lambda_i - 12E_{30}\right] - \\ - 16(K_4-K_6)\sum_i\lambda_i^3 + E_0^3\left[K_6 - 3K_2(K_4-K_2^2)\right] + \\ + 3E_0\left[2\mathrm{Sp}\,\mathbf{T}^2\cdot(2K_2-2K_4-K_2K_4+K_6) + 4(K_4+K_4K_2-K_2^2-K_6)\sum_i\lambda_i^2\right]. \quad (A7)$$

As it follows from Eqs.(A5) – (A7), the first three moments of the energy distribution are defined by the three first moments of the distribution of the quasienergies $\lambda_i$ (the energy $E_0$ is connected with the quasienergies by the relation $E_0 = -\sum \lambda_i$). Taking this into account and simplifying Eq.(A6) with the aid of Eq.(A1), we obtain the expression (8) for the variance $\sigma_n^2$.

For practical needs the most interesting is the form of the obtained expressions in the asymptotic limit $N \to \infty$. In this case the expressions (A1) transform into

$$K_1 = m, \qquad K_2 \approx m^2 - \frac{1-m^2}{N},$$

$$K_3 \approx m^3 - \frac{3m(1-m^2)}{N}, \qquad K_4 \approx m^4 - \frac{6m^2(1-m^2)}{N}, \qquad (A7)$$

$$K_5 \approx m^5 - \frac{10m^3(1-m^2)}{N}, \qquad K_6 \approx m^6 - \frac{15m^4(1-m^2)}{N},$$

where $m = 1 - 2n/N$ is the relative magnetization; we leave the terms $\sim O(N^{-1})$. Substituting these expressions in Eqs.(A5)-(A6), we obtain the asymptotic expressions (10) for the mean energy $\bar{E}_n$ and the variance $\sigma_n^2$. In this limit the expression for the third moment is defined by the asymptotic value of the quantity $R$:

$$R = 8(1-m^2)^3 \cdot \text{Sp}\mathbf{T}^3 - 16m^4(1-m^2)\sum_i \lambda_i^3 +$$

$$+ 2m^2(1-m^2)^2 \cdot \left[ 8\sum_{i,j} T_{ij}^3 s_i^{(0)} s_j^{(0)} - 24\sum_i (T^2)_{ii} \lambda_i - 12 E_{30} \right]. \qquad (A8)$$

Concluding the Appendix we present the exact expressions for the mean and the variance in the presence of the magnetic field

$$\bar{E}_n(\mathbf{H}) = K_2 E_0 - K_1 \mathbf{HS}_0^+,$$

$$\sigma_n^2(\mathbf{H}) = \sigma_n^2 + 4(K_1 - K_3) \cdot \mathbf{HTS}_0^+ - 2(K_3 - K_1 K_2) E_0 \cdot \mathbf{HS}_0^+ +$$
$$+ (1 - K_2) \cdot \|\mathbf{H}\|^2 + (K_2 - K_1^2) \cdot (\mathbf{HS}_0^+)^2 \qquad (A9)$$

where $\sigma_n^2$ is defined by Eq.(A6). In the asymptotic limit $N \to \infty$ these expressions take the form (12) and (13), respectively.